\documentclass[twocolumn,secnumarabic,amssymb, nobibnotes, aps, prd]{revtex4}
\usepackage{graphicx}

\begin{document}

\title{Chain of Superconducting Loops as a Possible Quantum Register}

\author{V.V.Aristov and A.V.Nikulov}
\email[]{nikulov@ipmt-hpm.ac.ru}
\affiliation{Institute of Microelectronics Technology and High Purity Materials, Russian Academy of Sciences, 142432 Chernogolovka, Moscow District, RUSSIA. }

\begin{abstract}
The idea of the quantum computation is based on paradoxical principles of quantum physics, superposition and entanglement of quantum states. This idea looks well-founded on the microscopic level in spite of the absence of an universally recognized interpretation of these paradoxical principles since they were corroborated over and over again by reliable experiments on the microscopic level. But the technology can not be able in the near future to work on the microscopic level. Therefore macroscopic quantum phenomenon - superconductivity is very attractive for the realization of the idea of quantum computer. It is shown in the present paper that a chain of superconducting loops can be only possible quantum register. The proposals by some authors to provide the EPR correlation with help of a classical interaction witness the misunderstanding of the entanglement essence. The problem of the possibility of superposition of macroscopically distinct states is considered.
\end{abstract}


\maketitle

\narrowtext

\section{INTRODUCTION}

The issue of quantum computation has attracted much attention in the last years \cite{Steane,Nielsen,Valiev,Kokin}. Quantum computer could perform certain tasks which no classical computer can perform in acceptable times. Some important computational tasks are impossible for any device apart from a quantum computer. Therefore the idea of quantum computation is very alluring. But there are very difficult problems on the way of its practical realization. The most obvious problem is technological difficulties.

The basic concepts of the quantum computation are quantum operations (gates) on two-state quantum systems called quantum bits or 'qubits' and register, i.e. array of entangled qubits. At first, only micro-systems, including cavity quantum electrodynamics \cite{photon}, ion and atom traps \cite{Cirac,atom}, nuclear spins \cite{spin} and others,  were proposed as qubits. But many experts are sure that to build a universal quantum computer on base of these proposals, i.e. on the microscopic level, is well beyond the abilities of current technology \cite{Steane}. Therefore macroscopic quantum phenomenon - superconductivity is very attractive for the realization of the idea of quantum computer.

The modern technology is able to make the mesoscopic superconducting circuits of ultra-small Josephson junctions which have been proposed as qubits \cite{Schoen97,Averin98,Mooij99S,Ioffe99}. But there are unsolved physical problems connected with obvious contradiction between quantum mechanics and macroscopic realism \cite{Legget85,Legget99,Legget02,Legget04}. Moreover there are philosophical problems since the idea of quantum computation is based on two paradoxical features of quantum physics: superposition of quantum states and entanglement. Many interpretations of these purely quantum phenomena were proposed \cite{Schreo35,Neuman55,Everet57,Zurek83,Bell87,Penros94,Zeili01E,Khren03L} but no interpretation is universally recognized up to now \cite{Mensky00,Lipkin,Nakhman,Pilan,Panov,Tsekhmis,Popov}. The problem of interpretation has not only philosophical but also practical importance. In order to propose possible realizations of quantum bits and registers it is needed to understand what is superposition and what is entanglement or even to understand what can not be superposition and entanglement.

One should not create illusions of understanding. Quantum mechanics contradicts to our everyday world experience and is not yet based on a generally accepted conceptual foundation \cite{Zeili99F}. Richard Feynman remarked: {\it I think I can safely say that nobody today understands quantum physics}. This remark may seem queer for many people but it is confirmed with the history of quantum physics, in particular, of the Einstein-Podolsky-Rosen (EPR) paradox. True, Richard Feynman said also that a habit may seem understanding. Some physicists confuse habit with understanding. It may be a most danger since habits may be bad.

But how can anybody propose quantum bits and quantum registers if nobody understand quantum physics? The sole reliable fulcrum is here reliable experimental results and a logical positivism. One should not have illusions that a formula deduced from others formulas describes without fail an reality but is not only an element of a science fiction \cite{fiction}. It is important first of all to understand what can not be entanglement and what can be reliable experimental evidence of superposition of macroscopic quantum states. Therefore the essence of entanglement is considered in the second section. A quantum register on base of superconductor structure is proposed in the third section and in the fourth section the problem of superposition of macroscopic quantum states is considered.

\section{THE ESSENCE OF ENTANGLEMENT}

From the two features of quantum physics, superposition and entanglement, the latter seems more important for the idea of quantum computation. In itself quantum superposition does not permit quantum computers to {\it perform many computations simultaneously} \cite{Steane00}. Although the entanglement is not possible without quantum superposition. The entanglement makes available types of computation process which, while not  exponentially larger than classical ones, are unavailable to classical systems \cite{Steane00}. Therefore it is important first of all to try to understand the essence of entanglement and its role in quantum computing.

The history of the idea of quantum computation may be useful for the elucidation of these problems. Many experts \cite{Steane} remark that the idea of quantum computation was provoked by Bell's 1964 analysis of the paradoxical thought-experiment proposed by Einstein, Podolsky and Rosen (EPR) in 1935 \cite{EPR}. The entanglement is called also Einstein- Podolsky- Rosen correlation. But it is important to emphasize that A. Einstein, B. Podolsky, and N. Rosen were fully confident that such correlation can not be. Therefore in order to understand what is not the entanglement the EPR work should be carefully analyzed.

\subsection{The Einstein-Podolsky-Rosen Paradox}
\label{sect:EPR}

It is no coincidence that J. S. Bell called his paper \cite{Bell64} "{\it On the Einstein-Podolsky-Rosen paradox}". Einstein, Podolsky and Rosen in  \cite{EPR} try to prove that the description of reality as given by a wave function is not complete using a paradoxical conclusion from a thought-experiment. They consider quantum systems consisting of two particles which interacted from the time $t=0$ to $t=T$, after which time EPR suppose that there is no longer any interaction between the two particles. This supposition by EPR seems very reasonable for the common sense when, for example, the particles are separated by some kilometers or even meters. EPR state also that the objective physical reality should exist with the criterion: {\it If, without in any way disturbing a system, we can predict with certainty (i.e., with probability equal to unity) the value of a physical quantity, then there exists an element of physical reality corresponding to this physical quantity}. On this basis of the supposition on the local realism EPR have proved that {\it the wave function does not provide a complete description of the physical reality}.

Indeed, according to the Heisenberg's uncertainty relation $\Delta p \Delta x \geq h$, one of the bases of the Copenhagen interpretation of quantum mechanics, when the momentum of a particle is known $\Delta p = 0$, its coordinate $\Delta x$ has no physical reality. The fundamental principle of the Copenhagen interpretation is the impossibility of noninvasive measurement. We can not measure precisely and simultaneously both momentum and coordinate since any measurement alters the state of quantum particle, a process known as the reduction of the wave function. But because of the law of conservation of momentum the measurement on momentum performed on, say, particle 1 immediately implies for particle 2 a precise momentum even when the two particles are separated by arbitrary distances without any actual interaction between them. Then, if the local realism is valid, i.e. the measurement performed on particle 1 can not alter the state of particle 2, we can define, contrary to the uncertainty relation, precise values both momentum and coordinate particle 2 after the measurement performed on its coordinate. Einstein, Podolsky and Rosen write in the end of the paper that one would not arrive at their conclusion if {\it the reality of momentum and coordinate of the particle 2 depend upon the process of measurement carried out on the particle 1, which does not disturb the state of the particle 2 in any way}. They state: {\it No reasonable definition of reality could be expected to permit this}.

Experiment has refuted even this statement on the local realism based on the common sense. But it is important to emphasize that strictly speaking this experiment has proved only the invalidity of the local realism but it does not prove completely the validity of the uncertainty relation. We could, following A. Einstein, B. Podolsky, and N. Rosen, ask: "Why can not we know with any exactness the momentum $p$ a particle measuring only its coordinates $x$ until the momentum $p =mv = mdx/dt$ is the product of mass $m$ and velocity $v =  dx/dt$ and therefore $\Delta p \Delta x = m\Delta v\Delta x = m(\Delta x)2/t < h$ at any uncertainty $\Delta x$ and an enough long time $t$?" There is not the assumption on the local realism and only the statement that the momentum is not the product of mass and velocity can save the absolute status of the uncertainty relation. In order the argumentation could be not merely circular this statement should be substantiated by experimental results but can not be base only on the Copenhagen interpretation of quantum mechanics. Until it is not proved experimentally that $p \neq mv = mdx/dt$ in all cases this interpretation seems not complete logically.

\subsection{The Bell's Inequality }

In 1964 John Bell \cite{Bell64}put the contradiction between the local realism and quantum mechanics into formulas.  He obtained certain bounds (Bell inequalities) on combinations of statistical correlations for measurements on two-particle systems if these correlations are understood within a realistic picture based on local hidden properties of each individual particle. In a realistic picture the measurement results are determined by properties the particles carry prior to and independent of observation. In a local picture the results obtained at one location are independent of any measurements or actions performed at space-like separation. Then Bell showed that quantum mechanics predicts violation of these constrains for certain statistical predictions for two-particle systems.

\subsection{Violation of Local Realistic Predictions}

By now a number of experiments \cite{Freedm72,Aspect81,Aspect82,Kwiat95,Pan00,Jennew02} have confirmed the quantum mechanical predictions contrary to the local realism. Already the first measurements of  the linear polarization correlation of the photons \cite{Freedm72,Aspect81} strongly violate the generalized Bell's inequalities, and rule out the realistic local theories. It is important that in accordance with the quantum mechanical prediction the correlation between results of the measurements does not depend on distance between the individual particles.

Already in 1981 \cite{Aspect81} no significant change in results was observed with source-polarizer separations of up to 6.5 m. The EPR correlation was observed on photons spatially separated by 400 m across the Innsbruck University science campus in \cite{Weihs98}, more than 10 km  in \cite{Tittel98,Zeilin04} and a Bell-experiment over thousands of kilometers is proposed \cite{Aspelm03,Kalten03}. Moreover the individual particles were truly space-like separated in the experiment by the Innsbruck team \cite{Weihs98}. A more striking conflict between quantum mechanical and local realistic predictions even for perfect correlations has been discovered for three and more particles, known as Greenberger-Horne-Zeilinger entanglement \cite{Bouwme99,Pan01}. Some other wonders of the entanglement are reviewed in \cite{Zeilin99}.

\subsection{Interpretations of Quantum Entanglement}

In contrast to the theories of relativity, entanglement and superpositions as well as a whole quantum mechanics are not yet based on a generally accepted conceptual foundation \cite{Zeili99F}. A number of coexisting interpretations utilizing mutually contradictory concepts \cite{Schreo35,Neuman55,Everet57,Zurek83,Bell87,Penros94,Zeili01E,Khren03L,Mensky00,Lipkin,Nakhman,Pilan,Panov,Tsekhmis,Popov}. The paradoxical Copenhagen concepts have engendered no less paradoxical interpretations. One of the less paradoxical point of view on quantum mechanics is the information-theoretical interpretation \cite{Zeili99F,Pilan,Zeili01E,Brukne02,Zeili96P}. This interpretation develops Schrodinger's ideas that the quantum entanglement is {\it entanglement of our knowledge}. It can explain some wonders of the entanglement but this method of approach is inclined to idealism.

Almost all interpretations of the entanglement have an inclination for idealism, though. Some interpretations \cite{Mensky00,Panov} follow the suggestion of Wigner \cite{Wigner} and others \cite{Pauli,Schrod44} that the observer's consciousness should be included in the theory of quantum measurement. But some authors \cite{Nakhman} state that we should not disturb the consciousness of observer because of the existence of microparticle consciousness. The author of \cite{Popov} defends quantum idealism. But under the circumstances no quantum idealism but realism  is needed in defence.

Possibly the coexistence of such a large number of philosophically quite different interpretations in itself contains an important message. One may suggest that the message is that a generally accepted foundational principle for quantum mechanics has not yet been identified and may agree with Richard Feynman that {\it nobody today understands quantum physics}.

\subsection{Entanglement is Purely Nonclassical Einstein-Podolsky-Rosen correlation}

But there is the striving for to reduce unclear empirical data to clear concept even if it is impossible. Many physicists think classically even when they have to do with quantum physics. It may be therefore the capacitive and inductive interactions were proposed \cite{Averin98,Mooij99S,Schoen99,Mooij99B,Schoen01,Freder03,Averin03} and even made \cite{Nakamu03,Mooij03,Iliche04} in order to couple superconductor qubits in quantum register. The  EPR correlation is purely non-classical phenomenon \cite{Bennet93}. The EPR experiment, in the form as analyzed by Bell, emphasizes that entanglement leads to a degree of correlation beyond that which can be explained in terms of local hidden variables \cite{Steane00}. Therefore it is strange that so many people can think that quantum register can be made on base of pure classical interactions.

The entanglement differs qualitatively from any classical interaction. According to conventional logic used in classical physics a system of $N$ degrees-of-freedom, each of their is described by only independent variable, is described by $N$ independent variables. But the quantum register of $N$ qubits is described by $2^{N} - 1$ independent variables. This advantage of quantum computer is a consequence of paradoxical quantum physics and can not be provided by any classical interaction. The entanglement between parts of a system takes place when the description of the system can not be reduced to the description of its parts. We could say that the parts of the system should described by a common wave function  but we can not imagine the wave function which can describe the common state of two photons separated by more than ten kilometers. It is important to note one again that according to quantum mechanics and the experiments \cite{Freedm72,Aspect81,Aspect82,Kwiat95,Pan00,Jennew02,Weihs98,Tittel98,Zeilin04} the entanglement takes place regardless of time and space. Bell emphasized in 1987 \cite{Bell87} that . . .{\it more importance, in my opinion, is the complete absence of the vital time factor in existing experiments}.

\section{HOW CAN THE ENTANGLEMENT BE MADE IN SUPERCONDUCTOR STRUCTURES?}
\label{sect:sections}

Niels Bohr wrote: {\it There is no quantum world. There is only an abstract quantum physical description. It is wrong to think that  the task of physics is to find out how Nature is. Physics concerns what we can say about Nature} \cite{Brukne02}. But we can say anything only on base of reliable experimental data. It is more difficult to propose quantum register on the macroscopic level than on the microscopic one since the reliable experimental evidences of the entanglement were observed on the microscopic level for the present. Only guide can be here the strangeness of entanglement and superconductivity.

\subsection{Strangenesses of Superconductivity}

One of the  strangenesses of superconductivity is such well known  phenomenon as the persistent current existing because of the quantization of the momentum circulation \cite{tink75}
$$\oint_{l}dl p =  \oint_{l}dl (mv + 2eA) = m\oint_{l}dlv + 2e\Phi = n2\pi \hbar \eqno{(1)} $$
This quantization is the cause of the Meissner effect $\Phi = 0$ and the quantization of the magnetic flux $\Phi = n2\pi \hbar/2e = n\Phi_{0}$ in the case of the strong screening and the velocity quantization
$$\oint_{l}dl v_{s} = \frac{2\pi \hbar}{m} (n -\frac{\Phi}{\Phi_{0}}) \eqno{(2)} $$
in the case of weak screening. These effects may be interpreted as a manifestation of an action which can be at any distance. According to the universally recognized point of view superconducting pairs are condensed bosons which are described by a common wave function $\Psi(r) = |\Psi|\exp(i\varphi )$ in a whole superconductor irrespective of its sizes. The integer number $n$ should equal zero $n = 0$ if the wave function $\Psi(r) = |\Psi|\exp(i\varphi )$ has not a singularity inside the $l$ path since $p = \hbar \nabla \varphi $ and $\oint_{l}dl \nabla \varphi = 0$ for any function $\varphi $ without singularity.
The Meissner effect is the astonishing corroboration of this mathematics.

 Superconductivity is macroscopic quantum phenomenon since superconducting pairs have the same value of the momentum circulation $n$ (1). The energy difference between adjacent permitted states for single electron $E_{e}(n+1) - E_{e}(n) = p_{n+1}^{2}/2m - p_{n}^{2}/2m = (2\pi^{2} \hbar^{2}/l^{2}m)(2n + 1)$ corresponds to a very low temperature for a real length $l$, for example $2\pi^{2}\hbar^{2}/l^{2}m \simeq k_{B} \ 0.01 K$ for $l \simeq 3 \mu m$, whereas in a superconducting loop $E_{s.p.}(n+1) - E_{s.p.}(n) \approx N_{s}\pi^{2}\hbar^{2}/l^{2}2m \gg k_{B}T$ since the number of pairs $N_{s} = V_{s}n_{s}$ is very great even near the critical temperature $T \approx T_{c}$. Therefore the persistent current in normal metal \cite{permet90,permet01} and semiconductor \cite{persem93,persem01} mesoscopic loops was observed first in 1990 \cite{permet90} in twenty years after first prediction \cite{Kulik} whereas its experimental evidence in superconductor was obtained by Meissner and Ochsenfeld as far ago as 1933.  The first experimental evidence of the persistent current $T > T_{c}$ \cite{KulikSup} was obtained in 1962 \cite{little62}.

The discreteness of the permitted states spectrum is higher in superconductor with larger volume $V_{s}$ since $E_{s.p.}(n+1) - E_{s.p.}(n) \propto N_{s}/l^{2} = V_{s}n_{s}/l^{2}$. The energy difference between permitted states (1) of a loop $E_{s.p.}(n+1) - E_{s.p.}(n) \propto s/l$ can be high even at very long length of the circumference $l$. For example $E_{s.p.}(n+1) - E_{s.p.}(n) \approx k_{B} 60 \ K$ at $l = 10 \ m$, the pair density $n_{s} \approx  10^{28} \ m^{-3}$ typical for $T \ll T_{c}$ and enough small cross-section $s \simeq 1 \ \mu m^{2}$. The thermodynamic average value of the pair velocity $\overline{v} \propto \overline{n} -\Phi/\Phi_{0}$ when $E_{s.p.}(n+1) - E_{s.p.}(n) \gg k_{B}T$. Where the thermodynamic average value $\overline{n}$ of the quantum number $n$ is close to an integer number $n$ corresponding to the minimum $(n -\Phi/\Phi_{0})^{2}$, when the magnetic flux $\Phi$ inside $l$ is not close to $(n + 0.5)\Phi_{0}$. Consequently the persistent current with the density $j_{s} = 2en_{s}\overline{v_{s}} \propto  (\overline{n} -\Phi/\Phi_{0})$ can be observed at $T \ll 60 \ K$ even in very long loop with $l = 10 \ m$ and $s \simeq 1 \ \mu m^{2}$ when $n_{s} \approx  10^{28} \ m^{-3}$ along the whole of the loop.

The equilibrium velocity $\overline{v_{s}} = (2\pi \hbar /ml)(\overline{n} -\Phi/\Phi_{0}) \simeq 10^{-5} \ m/s$, the density $j_{p} = 2en_{s}\overline{v_{s}} \simeq 3 \ 10^{4} \ A/m^{2}$ and the persistent current $I_{p} = sj_{p} \approx 3 \ 10^{-8} \ A$ in this loop at $(n -\Phi/\Phi_{0}) = 1/4$ when the superconducting state is closed, i.e. $n_{s} =  10^{28} \ m^{-3}$ along the whole of the loop. But $\overline{v_{s}} = j_{p} =I_{p} = 0$ when the density of superconducting pairs $n_{s} = 0$  even in very short loop segment $l_{seg}$, for example $l_{seg} \approx 1 \ \mu m = 10^{-7}l$. Thus, there is a quantum correlation between the equilibrium velocity $\overline{v_{s}}$ and  the pairs density $n_{s}$ in different loop segments separated by a macroscopic space, for example $l/2 = 5 \ m$. Superconducting pair is braked, i.e. its velocity decreases down zero, because of the pure classical electric force $mdv_{s}/dt = 2eE = -2e\nabla V$, where $V(t) = R_{seg}I(t) = R_{seg}I_{p} \exp(-t/\tau _{RL})$ is the potential difference because of a non-zero resistance $R_{seg} > 0$ at $n_{s} = 0$ in the segment $l_{seg}$ and $I(t) \neq 0$ during the time of current relaxation $\tau _{RL} = L_{l}/R_{seg}$. Here $L_{l}$ is the inductance of the loop $l$. The opposite change from $\overline{v_{s}} = 0$ to $\overline{v_{s}} = (2\pi \hbar /ml)(\overline{n} -\Phi/\Phi_{0}) \neq 0$ takes place because of the quantization (1) and is not induced by any classical force \cite{Hirsch}.   This pure nonclassical phenomenon has experimental corroboration.

The potential difference average during a long time $\Theta $ can be not zero $\overline{V} = \overline{R_{seg}I(t)} = \Theta ^{-1}\int_{0}^{\Theta }dt R_{seg}I(t) \propto I_{p} \propto (\overline{n} -\Phi/\Phi_{0})$ when the $l_{seg}$ is switched between $n_{s} \neq 0$ and $n_{s} = 0$ with a frequency $\omega_{sw} = N_{sw}/\Theta $ since the persistent current $I_{p}$ has the same direction at $\Phi \neq n\Phi_{0}$ and $\Phi \neq (n+0.5) \Phi_{0}$. At $\omega_{sw} \ll 1/\tau _{RL}$ the dc potential difference $V_{dc}(\Phi/\Phi_{0}) = \overline{V} = L_{l}I_{p}N_{sw}/\Theta =  L_{l}I_{p} \omega _{sw}$. The quantum oscillations $V_{dc}(\Phi/\Phi_{0})$ were observed on segments of asymmetric superconducting loops \cite{Dubon03,Dubon02,QuOs1967}. The dc voltage $V_{dc} \neq 0$ can be observed on the superconducting segment since the acceleration in the electric field $\overline{dp/dt} =2e\overline{E} = -2eV_{dc}/(l - l_{seg})$ is compensated with the momentum change from $\oint_{l}dl p = 2e\Phi$ to $\oint_{l}dl p = 2\pi \hbar n$ because of the quantization (1) when the loop reverts to the closed superconducting state \cite{JLTP98,PRB01}: $\oint_{l} dl \Delta p_{\Theta }/\Theta = (2\pi \hbar n - 2e\Phi)N_{sw}/\Theta  =2\pi \hbar (n - \Phi/\Phi_{0})\omega _{sw}$ at $\omega_{sw} \ll 1/\tau _{RL}$.

There are some strangeness in the quantum oscillations $I_{p}(\Phi/\Phi_{0})$ and $V_{dc}(\Phi/\Phi_{0})$ phenomena. The change of the $I_{p}$ and $V_{dc}$ direction with the $\Phi/\Phi_{0}$ value, without an external vector factor, is experimental evidence of an intrinsic breach of clockwise - anti-clockwise and right-left symmetries \cite{FQMT04}. The dc voltage $V_{dc}(\Phi/\Phi_{0})$ is observed since the $v_{s}$ value in superconducting loop segment can change because of the $n_{s}$ change in other segment. The essence of this non-local interaction is not intelligible as well as the essence of entanglement. But it is follow from experiment that both phenomena are observed when quantum particles is not local. This absence of  locality, like the main strangeness of entanglement, gives hope of possibility of quantum register on base of superconductor structure.

\subsection{Why any Classical Interaction can not provide with Entanglement}

Any classical interaction can not provide with entanglement first of all since these phenomena are qualitatively different. The entanglement takes place regardless of time and space whereas any classical interaction, even between quantum systems, can be described by local variables. According to the base idea of the quantum computation a quantum system can be considered as possible quantum register if its description can not reduced to description of its parts. Therefore the proposals \cite{Averin98,Mooij99S,Schoen99,Mooij99B,Schoen01,Freder03,Averin03} to entangle superconducting qubits with help of capacitive or inductive interactions can not provide a quantum register since each part (each qubits) of this system can be described by local variables and then the system of $N$ qubits can be described by no more than $N$ independent variables. No classical interaction can violate the local realism. And the experimental results \cite{Nakamu03,Mooij03,Iliche04} demonstrate full agreement with the local realism in spite of the statement by some authors on experimental evidence for entangled states.

\subsection{Chain of Superconducting Loops with Phase Coherence}

There is a chance to violate the local realism only if superconductor qubits are coupled at least by Josephson junction. Only in this case each superconducting pair can, according to the universally recognized point of view, be "smeared" and phase coherence can be over whole system. This demand restricts the possibility of the superconductor quantum register which can be proposed.  It is not clear how the Josephson qubit based on charge degrees of freedom \cite{Nakamu99,Nakamu01,Nakam03F}. can be entangled and could be in principle entangled at this restriction. The quantum register on base of superconductor flux qubits \cite{Mooij99S,Mooij99B,Mooij99M,Mooij02P,Mooij03S,Mooij04B} seems more perspective. The state of a chain of connected superconducting loops is described by a common wave function like the chain of entangled atoms or ions proposed one of the first as the quantum register \cite{Cirac,atom}. The loop with half of magnetic flux quanta is like to an atom with a spin. Therefore the chain of such loops may be considered as a possible quantum register. The Josephson junctions and variations of magnetic flux in each loop can be used for manipulations of state superposition of qubits and coupling between they.

\section{COULD BE SUPERPOSITION OF MACROSCOPIC QUANTUM STATES?}

It is less difficult to implement the chain of entangled superconducting loops than the chain of entangled atoms or ions since superconductivity is macroscopic quantum phenomenon. On the one hand it is obvious advantage since the entanglement can be assumed in superconductor structure with macroscopic sizes. But on the other hand there is a fundamental problem: {\it Could the superposition of states be in macroscopic quantum systems, such as superconductor structure, as well as it is observed in micro-systems?} Although enough many authors declared on experimental evidence for a coherent superposition of macroscopically distinct flux states \cite{QuanSup1,QuanSup2} the possibility of it is not clear for the present. There is important to understand a logical contradiction between quantum mechanics and macroscopic realism. The authors of \cite{Mooij03E} write: {\it In 1980, Leggett pointed out that cryogenic and microfabrication technologies had advanced to a level where macroscopic Schrodinger cat states could possibly be realized in small superconducting loops that contain Josephson tunnel junctions}. But it is important to note that the Schrodinger cat is that which is not possible in principle (certainly macroscopic one since nobody think to see a microscopic cat). The Leggett's papers, for example \cite{Legget85} with the significant title {\it Quantum mechanics versus macroscopic realism: Is the flux there when nobody looks?}, note that the Schrodinger challenge to the Copenhagen interpretation is not only merely philosophical problem but it can be tested now in experiment. In order to emphasize the fundamental nature of this problem and to confirm that Richard Feynman was right saying that {\it nobody understands quantum physics} we would like to show that a possibility of the coherent superposition of macroscopic state is less obvious than violation of the second law of thermodynamics.

\subsection{Violation of the Second Law of Thermodynamics is Most Ordinary and Obvious Consequence of Quantum Mechanics}

The contradiction with the second law is already in the well known and reliable experimental results \cite{permet90,permet01,persem93,persem01,little62}, first of they \cite{little62}was obtained as long ago as 1962. The observation of the quantum oscillations of the resistance $R(\Phi/\Phi_{0})$ of superconducting  loop \cite{Moshch} is experimental evidence of the persistent current $I_{p} \neq 0$, i.e. the equilibrium direct current,  observed at non-zero resistance $R > 0$ and consequently of a persistent power, i.e. an equilibrium dc power $RI_{p}^{2}$. Any dc power observed under equilibrium conditions is challenge to the second law since it is not random in contrast, for example, to the Nyquist's noise and therefore can be used for an useful work \cite{QI2002,VERHULST}.

There is a correlation between violation of the second law and an intrinsic breach of a symmetry \cite{FQMT04,VERHULST}. The equilibrium power of the Nyquist's noise can not be used since all elements of electric circuit have the same equilibrium frequency spectrum $W_{Nyq} = k_{B}T\Delta \omega $. There is a full symmetry. One can not say what element is power source and what one is load at $T_{1} = T_{2}$. This symmetry is broken when equilibrium conditions is broken $T_{1} > T_{2}$. One can distinguish a power source and load at $T_{1} = T_{2}$ only if their frequency spectrums are different, using a filter. But this difference can be only at an intrinsic breach of a symmetry.

Just such intrinsic breach of clockwise - anti-clockwise symmetry is observed in the persistent current phenomenon \cite{FQMT04}. It takes place because of discreteness of permitted state spectrum (1). Therefore the frequency spectrum the persistent power $RI_{p}^{2} \neq 0$ at $\omega = 0$ differs in essence from the Nyquist's noise spectrum $W_{Nyq} = k_{B}T\Delta \omega = 0$ at $\omega = 0$ and this quantum phenomenon is potential possibility of violation of the second law. The actual violation of the second law can be at the breach of right - left symmetry which is at the observation of the quantum oscillations of the dc voltage $V_{dc}(\Phi/\Phi_{0})$ \cite{Dubon03,Dubon02,QuOs1967}.

The persistent power $RI_{p}^{2}$ is fluctuation phenomenon like the Nyquist's noise. It is observed in a narrow fluctuation region near the superconducting transition $T_{c}$. Above this region at $T \gg T_{c}$, $R > 0$ but $I_{p} = 0$ and below $I_{p} \neq 0$ but $R = 0$ under equilibrium conditions. The persistent current $I_{p} \neq 0$ is observed at $R > 0$ since thermal fluctuations switch the loop between superconducting state with different connectivity, i.e. between the resistive $R > 0$ and superconducting $R = 0$, $I_{p} \neq 0$ states and therefore the dissipation force is compensated by the quantum force \cite{PRB01}.

\subsection{Fundamental Difference Between Quantum and Thermal Fluctuations}

The idea of flux (persistent current) qubit is based on the assumption that superposition of two eigenstates with opposite velocity, $v_{s}$ and $-v_{s}$, can be in superconducting loop at $\Phi = (n+0.5)\Phi_{0}$. According to the experimental results the $\overline{v_{s}^{2}}$ value is maximum \cite{little62,Moshch} whereas $\overline{v_{s}}=0$ \cite{Dubon03,Dubon02} at $\Phi = (n+0.5)\Phi_{0}$ in the fluctuation region near $T_{c}$. It takes place since the permitted states $n - \Phi/\Phi_{0} = 0.5$ and $n - \Phi/\Phi_{0} = -0.5$ have the same energy and consequently the same probability $P(v_{s}) = P(-v_{s})$: $\overline{v_{s}^{2}} = P(v_{s}) v_{s}^{2} + P(-v_{s}) (-v_{s})^{2} \neq 0$ and $\overline{v_{s}} = P(v_{s}) v_{s} + P(-v_{s}) (-v_{s}) = 0$. Therefore the persistent current $j_{p}(\Phi /\Phi_{0} ) = 2en_{s}\overline{v_{s}}$ equals zero not only at $\Phi = n\Phi_{0}$ but also at $\Phi = (n+0.5)\Phi_{0}$ near $T_{c}$ \cite{Dubon03,Dubon02}.

According to the theory \cite{Larkin02} the like dependence $j_{p}(\Phi /\Phi_{0} )$ can be observed at low temperature $T \ll  T_{c}$ because of quantum fluctuations. But it is not clear how the persistent current can be non-zero  $j_{p} \neq 0$ at $\Phi \neq (n+0.5)\Phi_{0}$ and zero $j_{p} = 0$ at $\Phi = (n+0.5)\Phi_{0}$ in the case of quantum fluctuation. The thermodynamic average $\overline{v_{s}} = 0$ at $T \approx T_{c}$ since the direction of the velocity changes in time because of thermal fluctuation. But it is impossible in the case of quantum fluctuation since the change in time of the current $dI_{p}/dt \neq 0$ should induce Faraday's voltage $\oint_{l}dlE_{F} = -d\Phi/dt = -LdI_{p}/dt \neq 0$ and as consequence an interchange of energy with environment.

\subsection{Quantum Mechanics Versus Macroscopic Realism}

The permitted states with $j_{p} \neq 0$ or $-j_{p} \neq 0$ can be observed at single measurement at $T \ll  T_{c}$ when the velocity $v_{s}$ direction can not change in time. According to the Copenhagen interpretation of quantum mechanics the loop can be in superposition of these states but only state can be observed. This is possible logically thanks to the principle of the impossibility of noninvasive measurement postulated by the Copenhagen interpretation. We are saved from the nightmare to observe the persistent currents flow at the same time in opposite directions only because of this principle if the superposition is possible. This principle seems admissible on the microscopic level when measuring device can not be smaller than measured object. But it becomes doubtful for the assumed case of superposition of macroscopic quantum states \cite{Ballen87}. We can not assume for the present that the Schrodinger cat can die or revive because of our look. Nevertheless some scientists have no doubt of the impossibility of noninvasive measurability even of macroscopic states.

\subsection{Experimental Results Obtained Far from Equilibrium can not be Evidence of Macroscopic Superposition}

Moreover some authors \cite{QuanSup1,QuanSup2} state on experimental evidence for coherent superposition of macroscopically distinct flux states and even on a detection of the Schrodinger cat. But the results \cite{QuanSup1,QuanSup2} obtained far from equilibrium can not bear a relation to the problem of the impossibility of noninvasive measurability. The statements by authors \cite{QuanSup1,QuanSup2} are based on the following logic: if an effect, like the one observed at superposition of microscopic quantum states, is observed in a macroscopic system then it is experimental evidence for superposition of macroscopical states. But this logic may be incorrect. Resembling effects can have different causes. The experimental results obtained for the present can not answer on the question in the title of \cite{Legget85}: {\it Is the Flux There When Nobody Looks?} And there is important question which should be resolved logically: {\it Can the flux $LI_{p}$ be change because of a look?} It is not clear what is the act of the "observation" reducing the superposition  $j_{p} \neq 0$ and $-j_{p} \neq 0$ to a single state, for example at measurement of $I_{p}$ by methods used in \cite{Geim} or \cite{Vanya}.

\subsection{Macroscopic Quantum Tunneling or Invisible External Noise}

The superposition of macroscopical states can not be without of a possibility of macroscopic quantum tunneling. Most scientists have no doubt of this possibility. Moreover a crossover observed in the temperature dependence of the escape probability out of the permitted state of superconducting loop is interpreted as experimental evidence for the macroscopic quantum tunneling \cite{MQT88,MQT02,MQT03}. The decrease of the escape probability with temperature decrease at high temperatures is interpreted as thermal activation regime and the stopping of this decrease at low temperature is interpreted as consequence of the macroscopic quantum tunneling. But the latter can be caused by an invisible unequilibrium noise which has in any measuring system. Very weak current impulse is needed in order to throw the superconducting state over the energy barrier with some degree height. When the unequilibrium noise is weaker than equilibrium one the escape probability decreases with the decrease of the latter, i.e. with the temperature decrease, as it is observed at $T > 0.6 K$ in \cite{MQT88} at $T > 0.3 K$ in \cite{MQT02} and at $T > 0.09 K$ in \cite{MQT03}. The decrease stops when the equilibrium noise becomes weaker at a low temperature than the unequilibrium one.

In order to be sure in any experimental evidence for the macroscopic quantum tunneling the power of the unequilibrium noise should be detected. Most suitable device for this purpose was proposed in \cite{detector} on base of a system of asymmetric superconductor loops connected in series. The switchings induced by equilibrium or nonequilibrium noise of the asymmetric loop with the persistent current are converted in the dc voltage \cite{Dubon03,Dubon02,QuOs1967} because of the intrinsic breach of the right - left symmetry \cite{FQMT04}. This dc voltage detects just the escape cause because of the noise. Its value and sign are the periodical function of the magnetic field $V(\Phi/\Phi_{0})$ like the persistent current and therefore this dc voltage can be easy picked out. Any how weak noise, right down to the equilibrium one, can be detected by the $V(\Phi/\Phi_{0})$ oscillations since the dc voltage is summed in loop system \cite{Dubon03} and the critical current, i.e. the energy barrier, can be reduced down to zero.

\section{Conclusions}

Although a possibility of a coherent superposition of macroscopically distinct flux states is less obvious than violation of the second law we think that the chain of superconducting loops is most perspective quantum register since there is little chance that the technology will be able in the near future to work on the microscopic level.

\acknowledgments

This work was financially supported by ITCS department of Russian Academy of Sciences in the Program "Technology Basis of New Computing Methods", by Russian Foundation of Basic Research (Grant 04-02-17068) and by the Presidium of Russian Academy of Sciences in the Program "Low-Dimensional Quantum Structures".

\end{document}